\documentclass[twocolumn]{aastex7}
\usepackage{hyperref}
\usepackage{amsmath}

\begin{document}

\title{Polar Filaments Capture High Latitude Solar Poloidal Field Interactions and can Foretell the Future Sunspot Cycle Amplitude before Polar Field Precursors}

\correspondingauthor{Dipankar Banerjee}
\email{dipu@iiap.res.in}

\author[0009-0008-5834-4590]{Srinjana Routh}
\affiliation{Aryabhatta Research Institute of Observational Sciences, Nainital-263002, Uttarakhand, India}
\affiliation{Department of Applied Physics, Mahatma Jyotiba Phule Rohilkhand University, Bareilly-243006, Uttar Pradesh, India }
\email{srinjana.routh@gmail.com}

\author[0000-0002-6130-7829]{Shaonwita Pal}
\affiliation{Center of Excellence in Space Sciences India, Indian Institute of Science Education and Research Kolkata, Mohanpur 71246, West Bengal, India}
\affiliation{Department of Technical Education, Training and Skill Development, Government of West Bengal, Newtown Rajarhat 700160, West Bengal, India}
\email{shao.physics@gmail.com}

\author[0000-0001-5205-2302]{Dibyendu Nandy}
\affiliation{Center of Excellence in Space Sciences India, Indian Institute of Science Education and Research Kolkata, Mohanpur 71246, West Bengal, India}
\affiliation{Department of Physical Sciences, Institute of Science Education and Research Kolkata, Mohanpur 71246, West Bengal, India}
\email{dibyendu.nandi@gmail.com}

\author[0000-0002-5014-7022]{Subhamoy Chatterjee}
\affiliation{Southwest Research Institute, Boulder, CO 80302, USA}
\email{fakeemail1@google.com}

\author[0000-0003-4653-6823]{Dipankar Banerjee}
\affiliation{Indian Institute of Space Science and technology, Valiamala, Thiruvananthapuram - 695 547
Kerala, India}
\affiliation{Indian Institute of Astrophysics, Koramangala, Bangalore 560034, India}
\affiliation{Center of Excellence in Space Sciences India, IISER Kolkata, Mohanpur 741246, West Bengal, India}
\email{dipu@iist.ac.in}

\author{Mohd. Saleem Khan}
\affiliation{Department of Applied Physics, Mahatma Jyotiba Phule Rohilkhand University, Bareilly-243006, Uttar Pradesh, India }
\email{subhamoy@boulder.swri.edu}

\begin{abstract}

Polar fields at the minimum of a sunspot cycle -- which are a manifestation of the radial component of the Sun's poloidal field -- are deemed to be the best indicator of the strength of the toroidal component, and hence the amplitude of the future sunspot cycle. However, the Sun's polar magnetic fields are difficult to constrain with ground-based or space-based observations from near the plane-of-ecliptic.  In this context, polar filaments -- dark, elongated structures that overlie polarity inversion lines -- are known to offer critical insights into solar polar field dynamics. Through investigations of the long-term evolution of polar filament areas and length acquired from the Meudon Observatory and complimentary solar surface flux transport simulations, here we establish the common physical foundation connecting the Babcock-Leighton solar dynamo mechanism of solar polar field reversal and build-up with the origin and evolution of polar filaments. We discover a new relationship connecting the residual filament area of adjacent solar cycles with the amplitude of the next sunspot cycle -- which can serve as a new tool for solar cycle forecasts -- advancing the forecast window to earlier than polar field based precursors. We conclude that polar filament properties encapsulate the physics of interaction of the poloidal magnetic field of the previous and current sunspot cycles, the resultant of which is the net poloidal magnetic field at the end of the current cycle, thus encoding as a precursor the strength of the upcoming solar cycle.
 
\end{abstract}


\section{Introduction} \label{sec:intro}

The Sun’s 11-year magnetic cycle manifests dramatically through a wide range of phenomena—from the regular rise and fall of sunspot numbers to the polarity reversal of the global magnetic field. Forecasting the strength and variability of future cycles remains one of the most critical challenges in solar physics, given its far-reaching implications for space weather and heliospheric dynamics \citep{Hathway2010, Petrovay2020, Nandy2021a, Nandy2023a}. Although surface proxies such as sunspots have long been used to trace and understand the cycle, attention has increasingly turned to more subtle indicators of solar activity, such as solar filaments, which evolve in tandem with deeper magnetic processes \citep{Li2010, Gao2012, Hao2013, Zou2014}. 

Solar filaments are dark, elongated features observed in H$\alpha$ on the solar disk, that form above polarity inversion lines (PILs) and act as tracers of the magnetic field's complexity \citep{Mcintosh1972,Martin1998}. Solar filaments, also known as "prominences" when observed at the limb, often appear as stick-like structures with barbs in the vicinity of which photospheric magnetic field cancellation occurs \citep{Tandberg1995,Chatterjee2017}. These features, though mostly extremely stable, are sometimes perturbed by hydrodynamic instabilities, leading to eruptive events like Coronal Mass Ejections (CMEs) and flares \citep{gilbert2000active,Gopalswamy2003,jing2004, Sinha2019,Majumder2023}. Due to their potential in aiding the understanding of the evolution of solar surface magnetic fields, filaments have been extracted from full-disk spectroheliograms as well as synoptic charts using both simple image processing methods and advanced machine learning algorithms \citep{Zharkova2005,Hao2015,Schuh2014,Chatterjee2017,webb2017,Priyadarshi2023}.

Solar filaments are traditionally divided into several types, namely, active region filaments, quiescent filaments, and polar filaments \citep{Mouradian1994}. The polar filaments, or the polar "crown" filaments \citep[PCFs; ][]{Xu2018,Xu2021}, form at the polarity inversion line of the large-scale magnetic field in high latitude \citep{Tlatov2016a,Tlatov2016b}. These structures merge at the intersection of fading magnetic flux from the previous cycle and the poleward-transported flux of the ongoing one \citep{Martin1998,diercke2019chromospheric}. Forming primarily during the rising phase of the cycle and often disappearing after the polar field reversal, these filaments occupy a transitional magnetic space that bridges the toroidal and poloidal components of the solar magnetic field. The origin and stability of these filaments are deeply rooted in the surrounding magnetic environment. Studies have consistently shown that flux convergence and cancellation are critical for filament formation, while their longevity is maintained by a delicate balance between the underlying PILs, coronal arcades, and magnetic helicity \citep{Mackay2010,martens2001origin}. 

Polar filaments exhibit a "rush-to-pole" phenomenon, which appears to be an indicator of polarity reversal and, thereby, the epoch of solar maximum \citep{Ananthakrishnan1952,Cliver2014}; these and several other studies collectively suggest that filaments manifesting at the poles align well with the timing of global solar polarity reversal \citep{Chatterjee2017, Mazumder2018, Mazumder2021}. We may then expect that high-latitude filaments serve as tracers of the large-scale evolution of surface magnetic flux in the polar regions and may reflect the very processes governing polar field dynamics. The polar fields of the Sun are important from the perspective of the Sun's global dipole field variation and the solar dynamo \citep{Jaswal2023}, high latitude coronal structure, open solar flux distribution, and cosmic ray modulation in the heliosphere \citep{Nandy2018,Pal2020,Dash2020,Dash2023}. However, constraining the Sun's polar magnetic fields remain  challenging and a top priority in solar and space physics \citep{Nandy2023b}.

In this Letter, through complementary polar filament observations and solar surface magnetic flux transport simulations, we demonstrate that polar filaments can serve as tracers of polar field reversal dynamics mediated via the emergence and dispersal of the flux of tilted bipolar active regions -- i.e., the so called Babcock-Leighton solar dynamo mechanism \citep{Kumar2019}. Furthermore, we discover a surprising correlation between polar filaments of past cycles and the future sunspot cycle that can act as a predictor of the strength of the future solar cycle. 

\begin{figure*}
    \centering
    \includegraphics[width=0.49\linewidth]{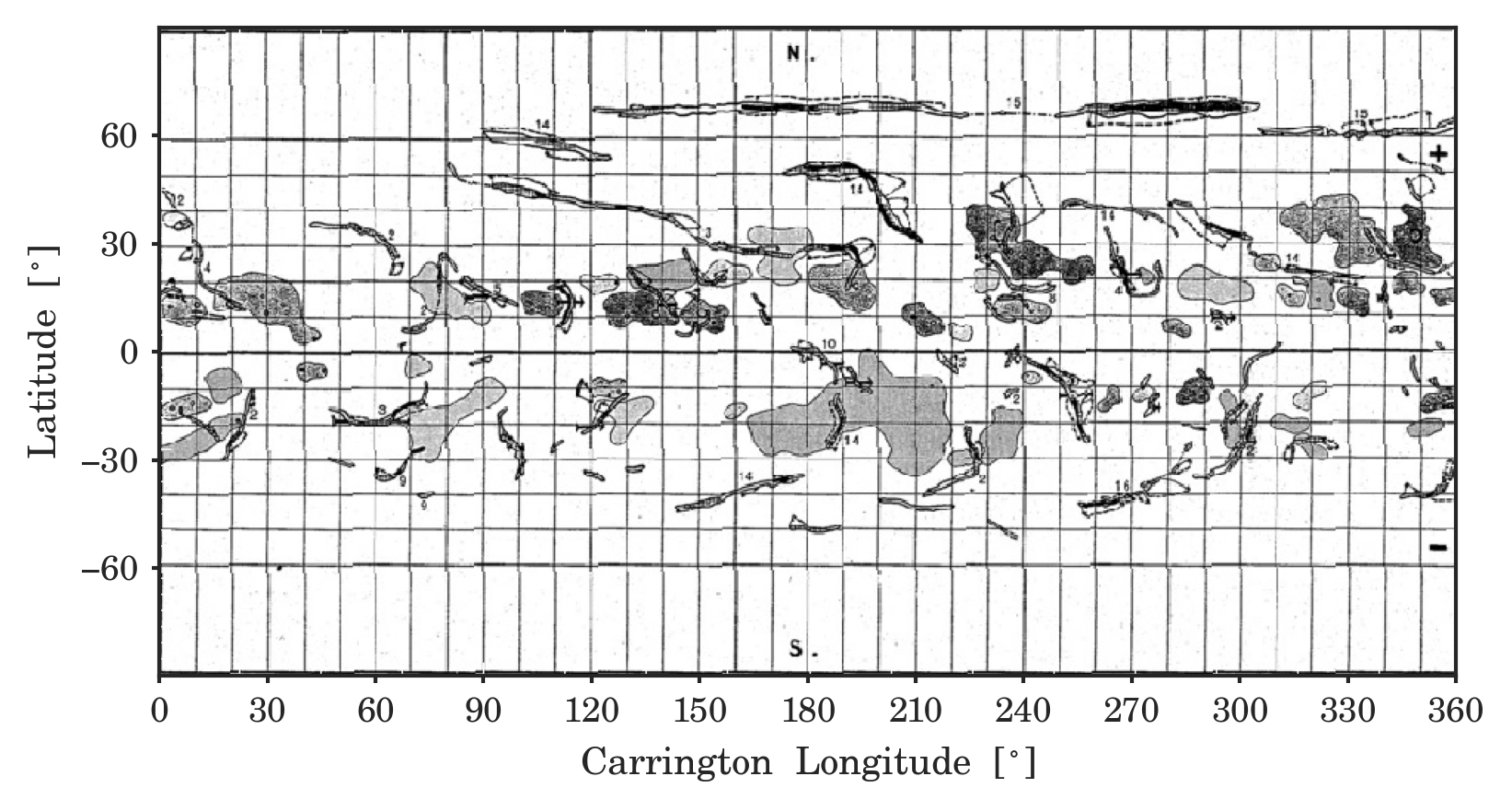} \includegraphics[width=0.49\linewidth]{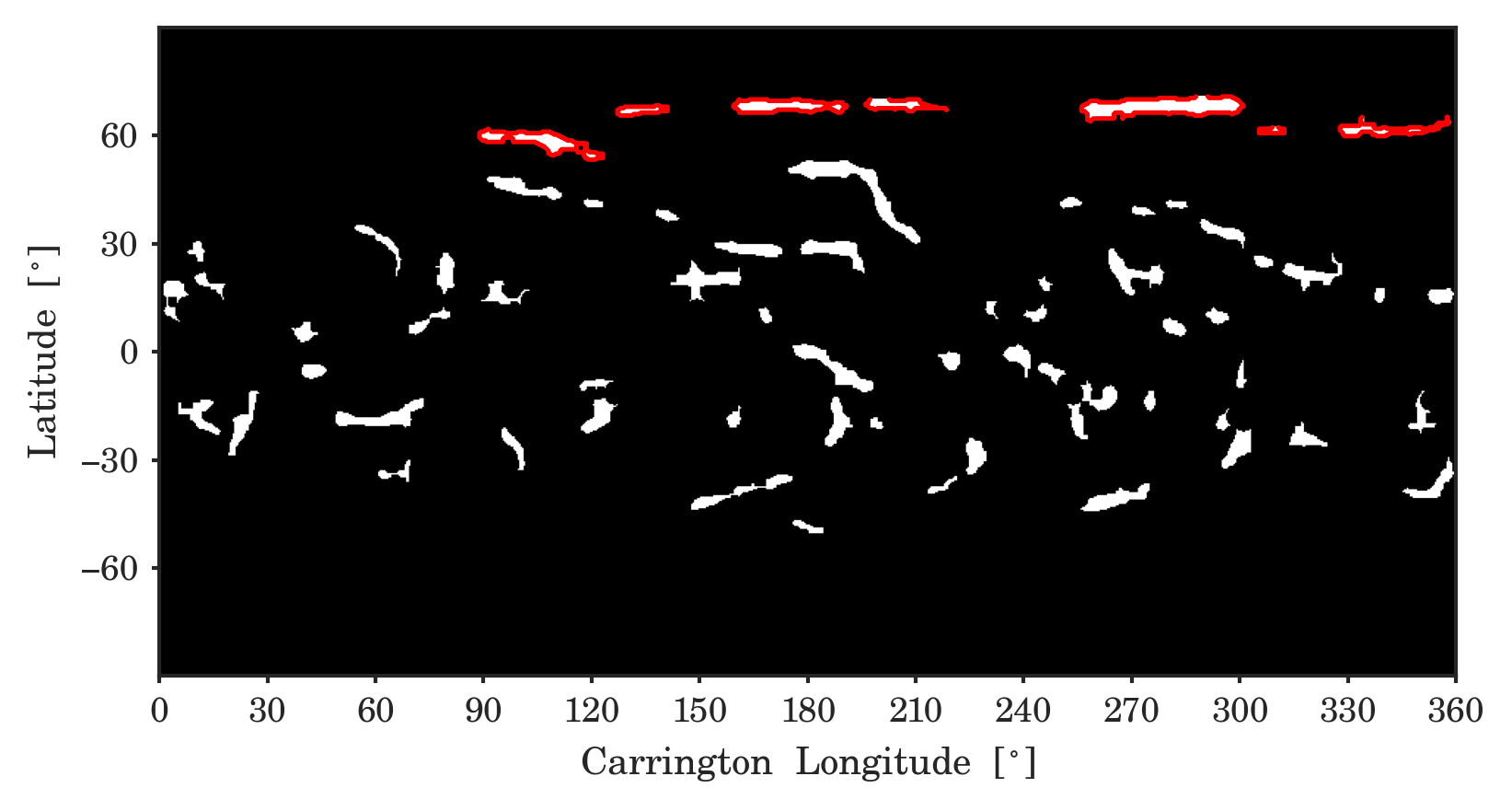}
    \caption{(Left Panel) A representative Meudon hand-drawn  synoptic chart after calibration corresponding to the Carrington rotation 1123. (Right Panel) A binary segmented image showcasing the extracted filaments, while the polar filaments ($|\theta|\geq50^{\circ}$) are separately identified with a red contour from the same.}
    \label{fig:calibrated_and_fil_detected}
\end{figure*}

\begin{figure*}
    \centering
    \includegraphics[width=0.49\textwidth]{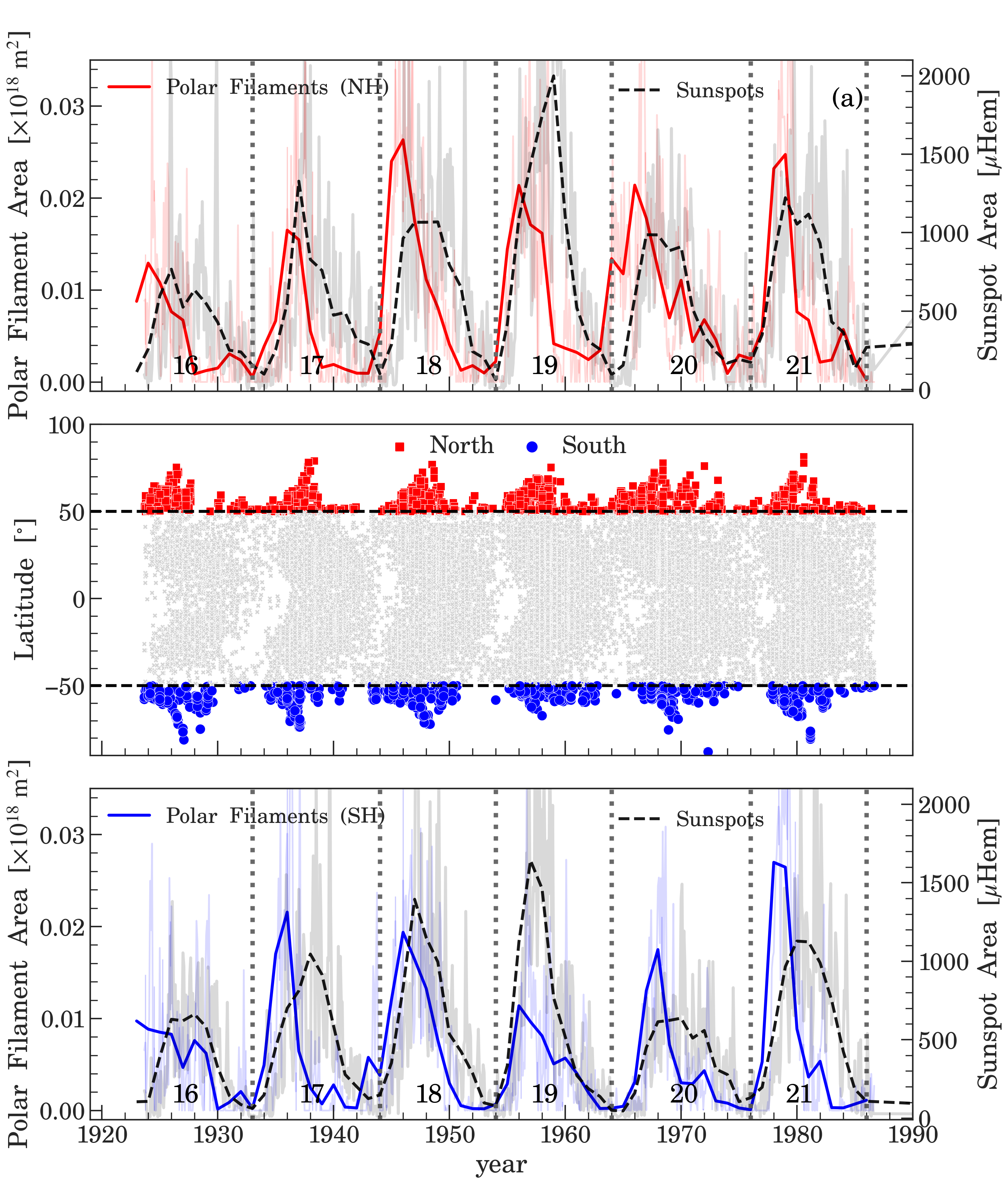} \includegraphics[width=0.49\textwidth]{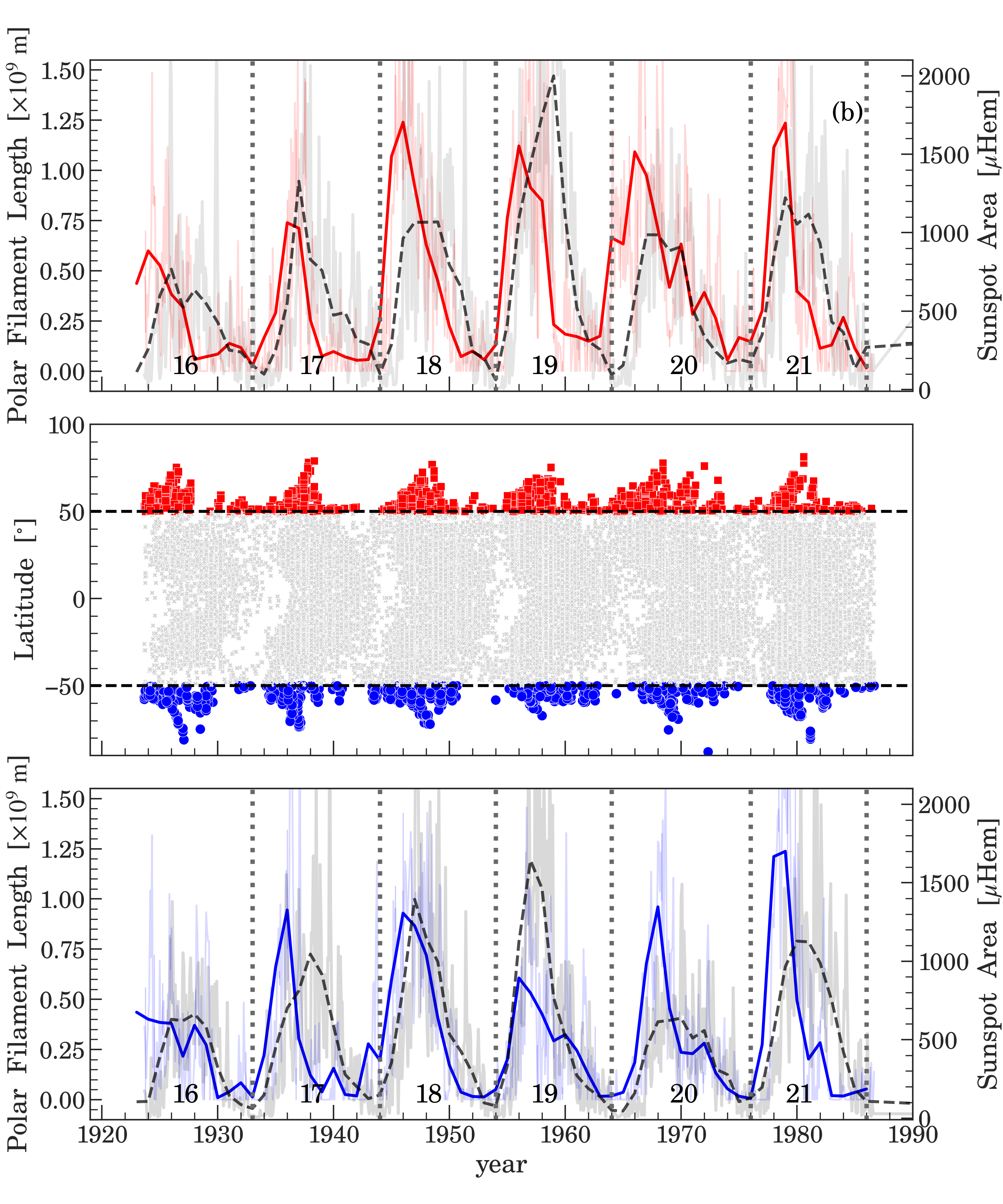}
    \caption{ Latitudinal distribution and temporal evolution of polar filament 
 (a) area and (b) length, marked by hemisphere for solar cycles 16–21. (Middle Panels) The latitudinal distribution of all filaments extracted is presented, revealing a characteristic butterfly-like pattern, with the polar filaments ($|\theta|\geq 50^{\circ}$) highlighted in red and blue for the North and South hemispheres, respectively. (Top Panels) The temporal evolution of (a) area and (b) length of the polar filaments in the northern hemisphere (shown in red continuous lines), compared with the sunspot area (shown in dashed black lines) in the same. The comparatively lighter red and black curves in the background show carrington rotation aggregated and monthly averaged polar filament parameters and sunspot area, respectively. Dimgrey dotted lines indicate the end epoch of each cycle. (Bottom Panels) The same in the southern hemisphere. All time series have been smoothed using a 13-month running average to enhance clarity. A coloured version of this diagram is available for the online version of this article.}
    \label{fig:butterfly}
\end{figure*}
\section{Data \& Methodology}\label{sec:data}

\subsection{Meudon Synoptic Maps}
The Meudon Observatory in France has regularly produced spectroheliograms in K1v, K3 and H$\alpha$ since 1919 \citep{Mouradian1998}, having a spatial resolution of $2-3$ arcseconds. The spectroheliograms are accompanied by hand-drawn synoptic charts\footnote{\url{https://bass2000.obspm.fr/lastsynmap.php}} since Carrington rotation 876 and have spanned cycles 15 to 23. These maps provide a synthetic representation of solar activity, including filaments, sunspots, and plages, based on daily spectroheliograms taken in the K1v, H$\alpha$, and K3 lines. The maps are drawn in cylindrical coordinates and represent each structure at its maximum development during its passage across the solar disk. Synthesized filaments were represented on the maps by two parallel lines, with a code indicating the daily visibility of the filament. The portion where the filament was visible for more than six days was marked in black, while the portion visible for one or two days was marked in white. Intermediate periods of visibility (three to five days) were indicated by hatching.

\subsection{Filament Detection}\label{subsec:fil_detect}

This study uses calibrated grayscale synoptic maps from \cite{Mazumder2021}\footnote{\url{https://github.com/rakesh-mazumder/calibrated-meudon-maps}}, derived from Meudon Observatory data, spanning Carrington rotations 876–1823 (grayscale) and 1824-2008 (color-coded). These maps fully cover solar cycles 16–21, while cycles 15, 22, and 23 are only partially included, and thus excluded from cycle-related analysis. To isolate filaments, a method similar to \cite{Mazumder2021} is used, with modifications better suited to extract polar filaments. A detailed discussion of this method is available in \cite{Mazumder2021}, however, the method is briefly summarized as follows,

\begin{enumerate}
    \item A denoising is performed using a 5 × 5 median filter, and pixels above an intensity of 160 are set to 255 to retain low intensity regions, including filaments.
    \item We perform a grayscale morphological opening on the resultant image from Step 1 to join the hatched patterns inside filaments and create continuous structures.
    \item We apply an intensity threshold of (median – $1.5\sigma$), producing a segmented image of filaments and plages. Due to hatching and shading-based representations of filaments encoding the information of their visibility, the method favors fully hatched/shaded filaments over those with only contour lines.
    \item A morphological opening operation with an elliptical kernel isolates plages and preserves the original morphology of the same, which is then subtracted from the image obtained from Step 3. This yields a segmented image that contains small-scale structures and filaments.
    \item Next, a connected component analysis removes small-scale structures under 20 pixels. A latitude threshold is then applied to the centroid of each filament to identify polar filaments. (see \autoref{fig:calibrated_and_fil_detected},right panel). 
\end{enumerate}

\subsection{Extraction of Polar Filament Parameters}\label{subsec:extract_fil_par}

Polar filaments from here on are defined using a latitude criterion, $|\theta|\geq50^{\circ}$. Although previous studies have defined "polar" filaments as those located at latitudes $|\theta|\geq45^{\circ}$ \citep{Leroy1983,Rust2000,Xu2018}, we opted for a higher threshold to eliminate any potential interference from active regions. An even higher threshold of ($|\theta|\geq55^{\circ}$) could have been implemented; however, this would have significantly reduced the available data, potentially compromising the statistics of our analysis. The length $l$ and area $a$ of the individual filaments were determined using the following formulae \citep{Mazumder2018,Mazumder2021},
\begin{equation*}
    l = \sum_{i} \sqrt{R_{\odot}^{2}\delta\theta^{2}+R_{\odot}^{2}\cos^{2}\theta_{i}\delta\phi^{2}}
    \label{eq:len}
\end{equation*}
\begin{equation*}
    a = \sum_{i}R_{\odot}^{2}\cos\theta_{i}\delta\theta\delta\phi
    \label{eq:area}
\end{equation*}
%With this threshold, $2,592$ polar filaments were identified among $29,853$ filaments extracted from $943$ maps.  
where, $R_{\odot}$ is the solar radius, $\theta$ and $\phi$ represent the latitude and longitude of a particular pixel $i$, respectively, and $\delta\theta$ and $\delta\phi$ are the longitudinal differences between two adjacent pixels in the latitudinal direction and longitudinal direction, respectively, which in this case amounts to the scale of the pixels in the $\theta$ and $\phi$ directions, respectively. These quantities are summed over $j$ individual filaments for a particular Carrington rotation to obtain rotation-integrated filament length $L$ and filament area $A$. The errors in the estimation of these two integrated filament parameters for the 1 Carrington rotation are calculated as follows,

\begin{align*}\label{eq:errors}
\centering
     \Delta L &= \frac{R_{\odot}\delta\theta\delta\phi}{2}\sqrt{\sum_{j}\left( L_{j}^{2}\sum_{i}\left(\frac{\sin^{2}2\theta_{ij}}{\cos^{2}\theta_{ij}\delta\phi^{2}+\delta\theta^{2}}\right)\right)}\\
     \Delta A &= R_{\odot}^{2}\delta\theta^{2}\phi^{2}\sqrt{\sum_{j}(\sum_{i} \sin^{2}\theta_{i})}   
\end{align*}

where, the quantities are summed over $i$ pixels for each of $j$ individual filaments. It is worth noting here that the errors obtained are an order less than the quantities, thereby rendering them barely notable against the large-scale variation seen across solar cycle (see \autoref{fig:butterfly}).

\subsection{Simulations of Historical Sunspot Cycles 16–21}\label{subsec:2.3}

\begin{figure*}
\centering
    \includegraphics[width=\linewidth]{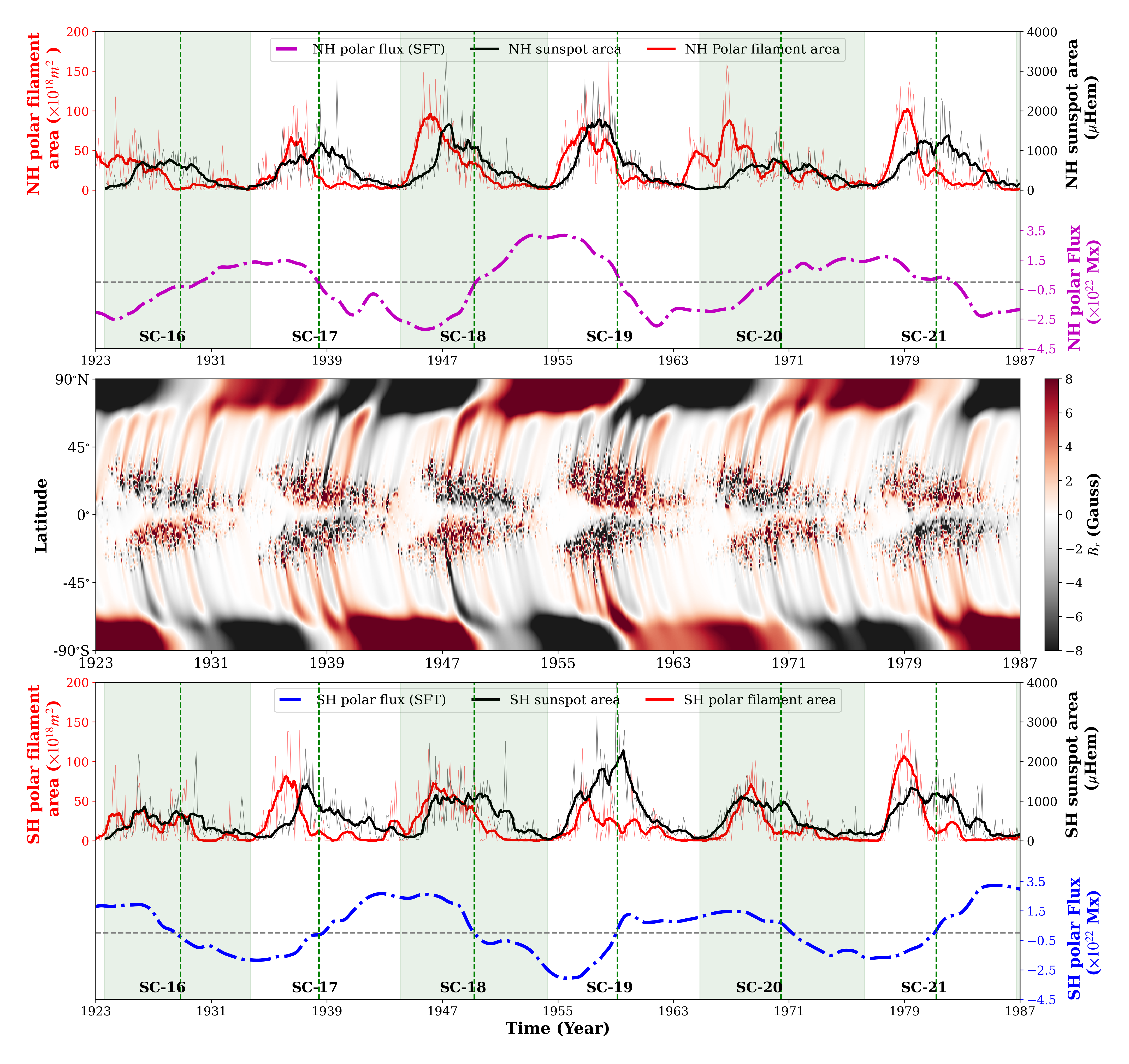}
    \caption{Top panel: The red curve shows the time series of polar filament area in the northern hemisphere from solar cycles 16 to 21, while the black curve represents the northern hemispheric sunspot area obtained from \cite{Mandal2020}. The lighter red and black curves in the background indicate the show carrington rotation aggregated and monthly averaged polar filament area and sunspot area, respectively, while the darker red and black curves show their corresponding 13-month running averages. The magenta curve denotes the northern hemispheric polar flux generated from data-driven optimized SFT model. Green dashed lines indicate the epochs of polar field reversal. Middle panel: This panel shows the spatio-temporal distribution of the radial magnetic field ($\mathrm{B_r}$) from solar cycles 16 to 21, generated using a data-driven optimized Surface Flux Transport (SFT) model (Pal and Nandy, in preparation). Red and black colors represent regions of positive and negative magnetic polarity, respectively. Bottom panel: The red curve shows the time series of polar filament area in the southern hemisphere from solar cycles 16 to 21, while the black curve represents the southern hemispheric sunspot area obtained from \cite{Mandal2020}. The lighter red and black curves in the background indicate show carrington rotation aggregated and monthly averaged polar filament parameters and sunspot area, respectively, while the darker red and black curves represent their corresponding 13-month running averages. The blue curve denotes the southern hemispheric polar flux generated from data-driven optimized. Green dashed lines indicate the epochs of polar field reversal.}
    \label{fig:3}
\end{figure*}

We utilize an observationally constrained, data-driven SFT model -- `SPhoTraM' \citep{Pal2025} -- to reconstruct surface magnetic field maps for sunspot cycles 16 to 21. This simulation is driven by observed sunspot emergence statistics, including latitude, longitude, area, and emergence time, obtained from the RGO/NOAA database. For sunspot tilt angles and polarity orientations, we employ a Monte Carlo ensemble approach, from which the best optimized realization is selected based on agreement with observational constraints. We compute the polar flux over the polar caps, defined by latitudes $\pm70^\circ$ to $\pm90^\circ$, in both hemispheres integrating the magnetic maps generated by SPhoTraM. These are shown in the first and third panels of \autoref{fig:3}. The SFT outputs are optimized by calibrating the resulting polar flux against observed polar faculae counts (see \cite{Pal2025} for methodological details and results). 

Importantly, this data-driven simulation framework enables the reconstruction of butterfly diagrams during periods lacking direct observations -- illustrated in the middle panel of the same figure \ref{fig:3} -- thereby offering a window into the historical evolution of solar magnetism.

\section{Results}\label{sec:res}

\subsection{Connection between the Babcock-Leighton Poloidal Field Generation Mechanism and Polar Filaments}\label{subsec:BLconnection}

The filaments serve as surface tracers of large-scale magnetic field evolution and offer valuable insights into the well-known Babcock-Leighton (BL) processes governing the buildup and reversal of the polar magnetic fields, which in turn, is a manifestation of the poloidal components of the magnetic field generated by the solar dynamo \citep{Cameron2015, Bhowmik2018, Charbonneau2020, Pal2024}. Sunspots that emerge on the solar surface eventually undergo decay and dispersal due to near-surface flows such as differential rotation, meridional circulation and turbulent diffusion. The net magnetic flux (corresponding to the sign of following polarity sunspots of bipolar sunspot pairs) is transported poleward by meridional flow, where it interacts with and cancels the pre-existing polar flux, ultimately generating new polar fields -- a process described by the BL mechanism \citep{Babcock1961, Leighton1969ApJ}. The transport of magnetic flux across the solar surface, driven by large-scale plasma flows, is effectively simulated using SFT models \citep{Jiang2014, Bhowmik2018, Pal2023, Yeates2023, Pal2024, Pal2025} as described in section \ref{subsec:2.3}. In this section, we utilize the polar flux and spatio-temporal magnetic field distribution generated by the SPhoTraM model to investigate how variations in polar filament area relate to the evolution of the Sun’s large-scale magnetic field.

A comparison of the filament parameters with the polar flux and radial component of the magnetic field is shown in \autoref{fig:3}, where the values of the polar flux and radial magnetic field are taken from \cite{Pal2025}. The bottom panel of \autoref{fig:3} illustrates the spatio-temporal evolution of the longitudinally averaged surface radial magnetic field ($\mathrm{B_r}$) from the optimized SFT simulation, covering solar cycles 16 to 21. This butterfly diagram depicts the poleward migration of the trailing polarity flux from decaying active regions from low-mid latitudes. This magnetic flux, which is opposite in sign to the pre-existing (old cycle) polar field, moves toward the polar regions where the opposite magnetic polarity interact. If this interaction -- mediated via the large-scale solar cycle dynamics due to the BL solar dynamo mechanism -- is the origin of polar filaments, then the imprint of this should manifest in the origin and variations of polar filaments associated with polarity inversion lines (PILs) in the interacting opposite polarity flux systems at high latitudes. We demonstrate that this is indeed the case.

In the top panel of \autoref{fig:3}, we observe that the polar filament area reaches its peak amplitude before the sunspot cycle maxima -- before the epochs of polar field reversal. This confirms that a majority of polar filaments form in the rising phase of solar cycles due to poleward transport and interaction of old and new cycle polar fields -- which leads to the cancellation of the old cycle fields and reversal in the polarity of the Sun's large-scale dipolar field. 

The reversal in the polar field occurs near solar maximum. Following this, the polar field starts building up, as evident in the SFT simulations. During this time, the polar surges of magnetic flux from lower latitudes have the same sign as the polar field, and thus, fewer filaments are expected. This is borne out in the observations, which show the filament area and length start dropping rapidly after solar maximum. However, we do find that there are still some filaments which appear at the declining phase of the cycle. This may result from poleward surges emanating from anomalous, anti-Hale active regions -- whose polarity orientation is opposite to the expected solar cycle trend \citep{Pal2023}. This comparative, long-term analysis of SFT simulations and polar filament dynamics suggests that polar filaments serve as effective proxies for the buildup of the large-scale polar field through the BL solar dynamo mechanism. 

\subsection{Polar filaments as an Indicator of Polar Flux Build-Up}\label{subsec:polarfieldvsssa}

Building on our earlier results in Section \ref{subsec:BLconnection}, which demonstrate a strong link between polar filament formation and polar field dynamics, we propose that polar filaments may themselves serve as potential precursors of upcoming cycle strength. To test this hypothesis, we consider that the flux transport processes underlying polar field dynamics are reflected in the physical characteristics of polar filaments -- such as filament length, $\mathrm{L}$, or filament area, $\mathrm{A}$. 

\begin{figure*}
\centering
    \includegraphics[width=\linewidth]{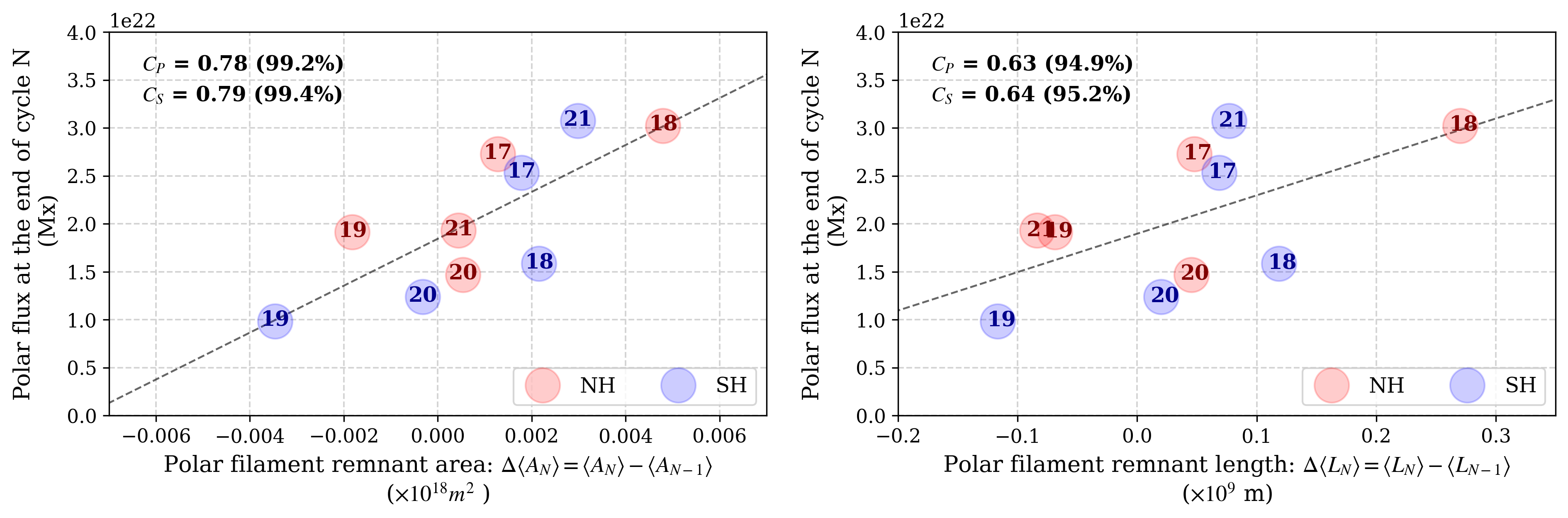}
    \caption{The left panel shows the correlation between the remnant average polar filament area, $\mathrm{ \Delta \langle A_N\rangle}$, and the polar flux at the end of the $\mathrm{N^{th}}$ cycle. The right panel displays the corresponding correlation between $\mathrm{ \Delta \langle L_N\rangle}$ and the polar flux strength of the end of sunspot cycle N. Red and blue circles represent data from the northern and southern hemispheres, respectively. The cycle number labeled at each circle refers to cycle $\mathrm{N}$. The black dashed line in both panels indicates the best linear fit to the data.}
    \label{fig:4}
\end{figure*}

We compute the average filament area $\langle \mathrm{A} \rangle$ and length $\langle \mathrm{L} \rangle$ for each solar cycle by dividing the total integrated filament area and length over the cycle by its duration. The differences between successive cycles are then calculated to define the remnant polar filament quantities: (1) the remnant filament area as $\mathrm{\Delta \langle A_N\rangle = \langle A_N\rangle - \langle A_{N-1}\rangle}$ and (2) the remnant filament length as $\mathrm{\Delta \langle L_N\rangle = \langle L_N\rangle - \langle L_{N-1}\rangle}$. To investigate whether polar filaments act as reliable indicators of the residual magnetic flux, we perform a statistical correlation analysis between the remnant filament area, $\mathrm{\Delta \langle A_N\rangle}$, and the polar flux amplitude at the end of sunspot cycle $\mathrm{N}$ \footnote{the simulated polar flux amplitude data are taken from \cite{Pal2025}}. We find a strong and statistically significant positive correlation between these two quantities (see the first panel of \autoref{fig:4}). A moderate-to-strong positive correlation is also found between the remnant filament length, $\mathrm{ \Delta \langle L_N\rangle}$, and the polar flux amplitude at the $\mathrm{N^{th}}$ cycle's end, although this correlation is somewhat weaker compared to that based on area, as shown in the second panel of \autoref{fig:4}). 

The presence of polar filaments traces the amount and complexity of remnant flux at high latitudes, where PILs form due to the interaction of opposite-polarity remnant flux. Now, the difference in average polar filament area between two cycles can be interpreted as a proxy for the net flux transported poleward during cycle N. A positive difference implies more filament material and hence more accumulated flux, which should lead to stronger polar flux build-up at the end of cycle N. 

\subsection{Polar Filaments as a Precursor for the Strength of Sunspot Cycles}\label{subsec:np1nvssa}

It is both observationally and theoretically established that the polar magnetic field at solar minimum correlates well with the strength of the subsequent solar cycle, making it a reliable predictor of future solar activity \citep{Yeates2008, Munoz2012, Nandy2021b}. Building on this, we compute the correlation between the remnant polar filament area, denoted as $\mathrm{ \Delta \langle A_N\rangle}$, and the peak sunspot area of the $\mathrm{N+1^{th}}$ solar cycle, which serves as a proxy for the toroidal magnetic field strength \footnote{The sunspot area data are taken from \cite{Mandal2020}}. A similar correlation analysis is also performed between the remnant filament length, $\mathrm{ \Delta \langle L_N\rangle}$, and the maximum sunspot area. 

\begin{figure*}
   \centering
    \includegraphics[width=\linewidth]{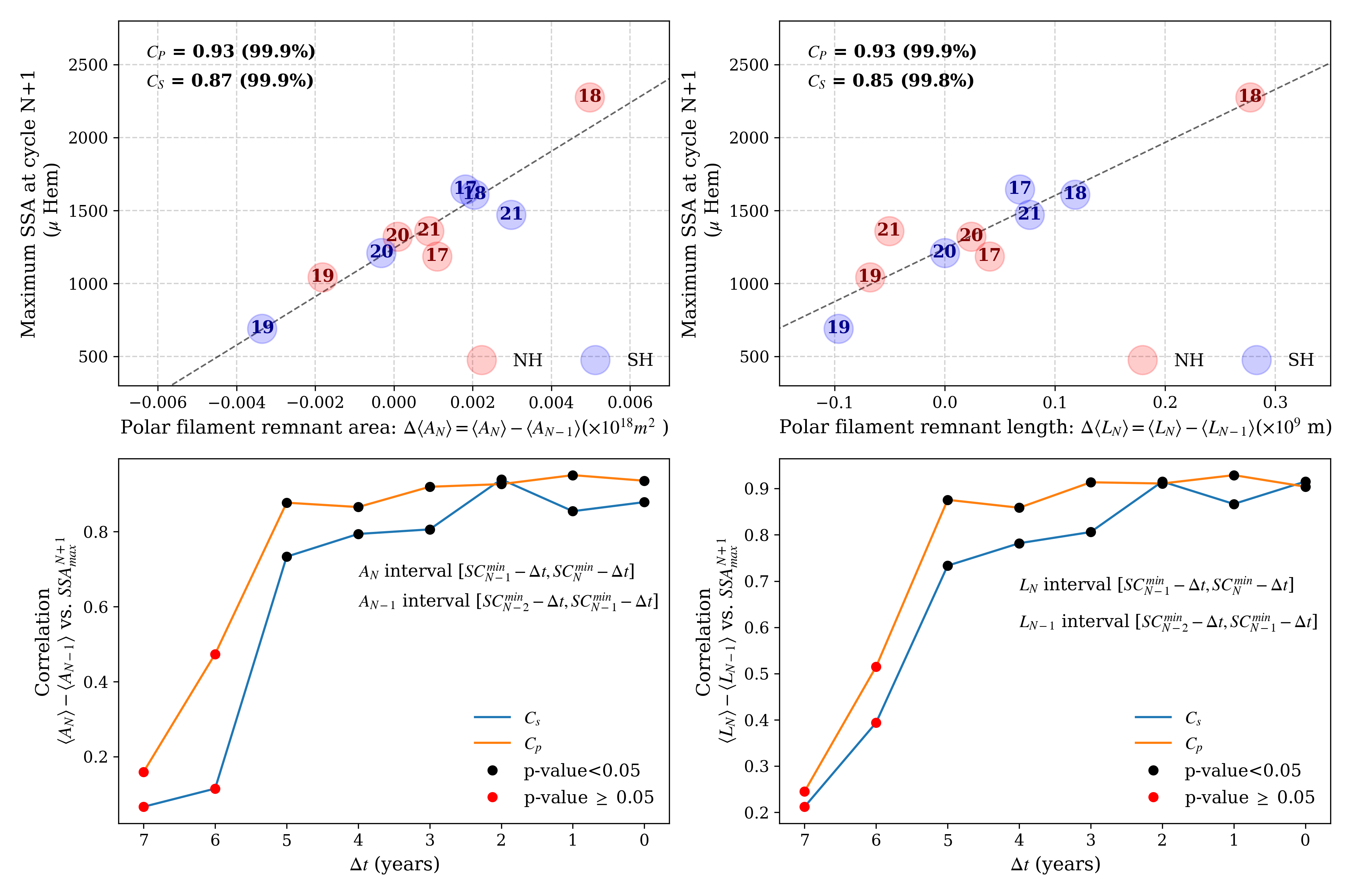}
    \caption{Correlation of polar filament remnant over solar cycle (SC) $N$ and $N-1$  with maximum sunspot area (SSA) of cycle $N+1$. The {top-left} panel shows the correlation between the remnant average polar filament area, $\mathrm{ \Delta \langle A_N\rangle}$, and the maximum sunspot area of the $\mathrm{{N+1}^{th}}$ cycle. The top-right panel displays the corresponding correlation between $\mathrm{ \Delta \langle L_N\rangle}$ and the peak sunspot area of the next cycle. Red and blue circles represent data from the northern and southern hemispheres, respectively. The cycle number labeled at each circle refers to cycle $\mathrm{N}$. The black dashed line in both panels indicates the best linear fit to the data. Bottom panels show the result from sliding the correlation coefficient in time $\Delta$t before the end of cycle N calculated for the $\mathrm{ \Delta \langle A_N\rangle}$ (left) and $\mathrm{ \Delta \langle L_N\rangle}$ (right), showing statistically significant (p-value$<$0.05) correlation with $SSA_{max}^{N+1}$ for $\Delta t \leq 5$ years measured relative to the solar minimum of cycle $N$.}
    \label{fig:5}
\end{figure*}

We find a strong and statistically significant positive correlation for both polar filament parameters, area and length, as shown in the first and second panels of \autoref{fig:5}. This result indicates that polar filament properties do serve as a proxy for polar field buildup and the amplitude of the next sunspot cycle and may serve as a new proxy for solar cycle predictions. 

To estimate how early this prediction can take place, a sliding correlation analysis is performed by calculating the correlation between polar filament proxies and the next cycle's peak sunspot area at various times ($\Delta$t) before the cycle ends. To estimate the correlation as a function of $\Delta t$, (1) we calculate the average polar filament area ($\mathrm{ \langle A_N\rangle}$) and length ($\mathrm{\langle L_{N}\rangle}$) from $\Delta$t years before start of cycle $N$ to $\Delta$t years before the end of cycle $N$ (2) we calculate the same ($\mathrm{\langle A_{N-1}\rangle}$ and $\mathrm{\langle L_{N-1}\rangle}$) for the cycle ($N-1$), and finally, (3) correlate the difference ($\mathrm{\Delta \langle A_N\rangle}$ and $\mathrm{\Delta \langle L_N\rangle}$) with the maximum sunspot area of cycle $N+1$. This entire process was repeated for different values of $\Delta$t, ranging from 0 to 7 years in steps of 1 year. The bottom panels in \autoref{fig:5} show that both $\mathrm{ \Delta \langle A_N\rangle}$ and $\mathrm{ \Delta \langle L_N\rangle}$) remain strongly correlated ($C_p\geq$ 0.88 (99.9\%), $C_s\geq$ 0.73 (98.4\%)) with future ($N+1$) sunspot area cycle strength even when measured up to $\Delta t$ of $\approx 5$ years. The correlation value as well as its statistical significance drops drastically for $\Delta t > 5$ years. This implies that our technique of polar filament based cycle prediction can advance empirical prediction window to about 10 years before the peak of a solar cycle. The loss in predictability beyond this window is consistent with the fact that once the previous maxima are crossed, we start missing a significant number of filaments that persist at least until the maxima.

% ($\rho\leq$ 0.33 (65.2\%), $\rho_{s}\leq$ (61.5\%) for area and $\rho\leq$ 0.27 (56.1\%), $\rho_{s}\leq$ 0.28 (58.1\%) for length).}

\section{Conclusion}\label{sec:conc}

In this study we explore the connection between polar filament dynamics and the solar magnetic cycle, utilizing a combination of Meudon hand-drawn synoptic charts and results from surface flux transport simulations. Our results demonstrate that polar filaments can serve as effective tracers of the solar magnetic field’s evolution, particularly during the polar field reversal and early build-up phases governed by the Babcock-Leighton dynamo mechanism. The formation and variability of polar filaments are closely linked to the poleward transport and interaction of oppositely signed magnetic flux, indicating that these features may represent observable signatures of the solar magnetic cycle.

We specifically examine whether polar filament properties can serve as proxies of polar field interactions between adjacent cycles, ($N-1$) and ($N$), which result in the net polar field at the end of the current cycle $N$. We discover a statistically significant and strong relationship between the residuals of the cycle-integrated filament properties (area and length) and the amplitude of the following sunspot cycle. 

Our findings suggest that polar filament properties capture the physics of interactions of the surviving (surface component of the) poloidal magnetic field from the previous cycle with the new poloidal field of the current cycle; the net effect of this interaction is the residual poloidal field at the end of the current cycle which determines the strength of the following sunspot cycle. This opens up the possibility of defining the cycle-averaged polar filament area or length as a new type of polar precursor, potentially enabling early prediction of future solar cycles, well before the minimum of the preceding cycle.

 We conclude that polar filaments show strong promise as supplementary indicators of solar cycle strength. This may serve as an opportunity for empirically predicting future sunspot cycle amplitudes earlier than the existing polar field proxies at solar minima. However, we reiterate that unlike physical model based predictions \citep{Bhowmik2018, Nandy2021b}, we cannot utlise this technique to predict the timing of the next sunspot cycle peak. 
We note that our study was performed using only one database and we anticipate further analysis with other databases to add to our findings. We also notice a reduction in the statistical confidence as well the correlation coefficients itself when the calculations are performed separately for each hemisphere. This may either be due to low cycle statistics or an intrinsic ability of cumulative, full-disk filament areas and lengths to more accurately reflect the large-scale solar dipole moment. We anticipate further investigations in this promising direction may establish this potential as a viable alternative to polar field diagnostics based solar cycle predictions.

\begin{acknowledgments}
S.R. is supported by funding from the Department of Science and Technology (DST), Government of India, through the Aryabhatta Research Institute of Observational Sciences (ARIES). The computational resources utilized in this study were provided by ARIES and the Center of Excellence in Space Sciences India (CESSI). CESSI is supported by IISER Kolkata, Ministry of Education, Government of India. The authors thank the staff at the Meudon Observatory for maintaining the observational data.
\end{acknowledgments}

\begin{contribution}
All authors have made substantial contributions to the work reported in this manuscript.

\end{contribution}
\software{scipy, numpy, OpenCV, sunpy, pandas, seaborn.}
\bibliography{references}{}

\bibliographystyle{aasjournal}

\end{document}